\documentclass[a4paper]{jpconf}

\usepackage{graphicx}
\usepackage{amsmath}
\usepackage{amssymb}
\usepackage{latexsym}

\begin{document}
\title{Magnetic behavior of a spin-1 Blume-Emery-Griffiths model}

\author{F. P. Mancini$^{1,2}$}

\address{$^1$ Dipartimento di Fisica {\it ``E. R. Caianiello"}
and Laboratorio Regionale SuperMat, CNR-INFM, Universit\`a degli
Studi di Salerno, Via S. Allende, I-84081 Baronissi (SA), Italy}
\address{$^2$ I.N.F.N. Sezione di Perugia, Via A. Pascoli, I-06123
Perugia, Italy}

\ead{fpmancini@sa.infn.it}

\begin{abstract}
I study the one-dimensional spin-1 Blume-Emery-Griffiths model
with bilinear and biquadratic exchange interactions and single-ion
crystal field under an applied magnetic field. This model can be
exactly mapped into a tight-binding Hubbard model - extended to
include intersite interactions - provided one renormalizes the
chemical and the on-site potentials, which become temperature
dependent.  After this transformation, I provide the exact
solution of the Blume-Emery-Griffiths model in one dimension by
means of the Green's functions and equations of motion formalism.
I investigate the magnetic variations of physical quantities -
such as magnetization, quadrupolar moment, susceptibility - for
different values of the interaction parameters and of the applied
field, focusing on the role played by the  biquadratic interaction in the
breakdown of the magnetization plateaus.
\end{abstract}

\section{Introduction}

The Blume-Emery-Griffiths (BEG) model is a spin-1 model which
presents a rich variety of critical and multicritical phenomena
\cite{blume71}. This model has been used to describe  systems
characterized by three states per spin, and it was originally
introduced to describe the phase separation and superfluidity in
the $^3$He-$^4$He mixtures. The BEG model can also describe the
properties of a variety of systems ranging from spin-1 magnets to
liquid crystal mixtures, semiconductor alloys, microemulsions, to
quote a few. The spin-1 BEG model is characterized by a bilinear
($J$) and biquadratic ($K$) nearest-neighbor pair interactions and
a single-ion potential ($\Delta$). The one-dimensional case for
the BEG model was studied in Ref.~\cite{krinsky75}, where exact
renormalization-group recursion relations were derived, exhibiting
tricritical and critical fixed points.

The underlying idea of this paper is to provide a new and general
framework to exactly investigate the one-dimensional BEG model in
the whole parameters space. For one-dimensional lattices, or more
generally for lattices with no closed loops, classical fermionic
and spin systems can be easily solved by means of the transfer
matrix method \cite{baxter}. As it is well known,  the partition
function of a 1d model of a classical spin-1 system with
nearest-neighbor interactions can be calculated from the largest
eigenvalue of a $3 \times 3$ matrix. However, this method is hardy
implementable when more complex lattices are considered. The
Onsager solution for the two-dimensional Ising model is an
emblematic example \cite{onsager}. Thus, there is the necessity to
foster alternative methods which can be used for  a large class of
lattices. Recently, it has been  shown that, upon transforming to
fermionic variables, spin systems can be conveniently studied by
means of quantum field methods, namely: Green's functions and
equations of motion methods \cite{mancini05s}. This approach has
the advantage of offering a general formulation for any dimension
and to provide a rigorous determination of a complete set of
eigenoperators of the Hamiltonian and, correspondingly, of the set
of elementary excitations.

%

%

%
The aim of the present paper is twofold. First, I would like to
further develop our previous work \cite{cmp_08}, where the finite
temperature phase diagrams of the one-dimensional BEG model with
vanishing biquadratic exchange have been obtained, by extending it
to a more general situation with a finite biquadratic interaction.
Secondly, the BEG model exhibits interesting features when one
considers a non-zero biquadratic coupling, such as the breakdown
of magnetic plateaus. Here, I address the problem of determining
the effect on the behavior of thermodynamic quantities of the
presence of a finite biquadratic interaction. I study the
properties of the system as functions of the external parameters
$J$, $T$, $\Delta$ and $h$ allowing for the biquadratic
interaction $K$ to be both repulsive and attractive.

\section{The model}
\label{sec_II}

The BEG model consists of a system with three states per spin. For
nearest-neighbor interaction, the one-dimensional BEG model is
described by the Hamiltonian
\begin{equation}
\label{eq1} H=-J\sum_\textbf{i} S(i)S(i+1)-K\sum_\textbf{i} S^2(i)S^2(i+1)+\Delta
\sum_\textbf{i} S^2(i)-h\sum_\textbf{i} S(i),
\end{equation}
where the spin variable $S(i)$ takes the values $S(i)=-1,0,1$. I
use the Heisenberg picture:  $i=(\textbf{i},t)$, where
$\textbf{i}$  stands for the lattice vector  $\bf R_i$.  This
model can be mapped, in two steps, into a fermionic model: namely,
the Hubbard model (in the limit of zero bandwidth) with local
interaction $U$ and extended to include intersite, charge-charge,
charge-double occupancy and double occupancy-double occupancy
interactions \cite{cmp_08}. Firstly, by means of the
transformation $S(i)=[n(i)-1]$, where $n(i)=\sum_\sigma c_\sigma
^\dag (i)c_\sigma (i)=c^\dag (i)c(i)$ is the density number
operator of a fermionic system. $c(i)$ ($c^\dag (i))$ is the
annihilation (creation) operator of fermionic field in the spinor
notation and satisfies canonical anti-commutation relations.
Secondly, by taking into account the four possible values of the
particle density ($n(i)=0$, $n_{\uparrow}(i)=1$ and
$n_{\downarrow}(i)=1$, $n(i)=2$), one redefines the chemical
potential as $\mu=\mu'-\beta ^{-1}\ln 2$ and the on-site potential
 as $U=U' -2\beta ^{-1}\ln 2$, where $\beta =1/k_B T$.
I refer the interested reader to Ref. \cite{cmp_08,joam} for a
detailed analysis of the mapping between the extended Hubbard
model and spin-1 Ising model. Once the BEG model has been mapped
into  the fermionic model, the latter can be exactly solved by
means of the Green's function and equations of motion formalism.
Upon introducing the Hubbard projection operators $\xi
(i)=[1-n(i)]c(i)$, and $\eta (i)=n(i)c(i)$, one can define
composite multiplet operators $\psi ^{(\xi )}(i)$ and $\psi
^{(\eta)}(i)$ which are eigenoperators of the fermionic
Hamiltonian  with eigenenergies $ \varepsilon ^{(\xi )}$ and $
\varepsilon ^{(\eta)}$ \cite{cmp_08}. As a consequence, an exact
solution of the Hamiltonian can be formally obtained. Within this
framework, the retarded Green's function  and pertinent
correlation functions turn out to be functions of only two
parameters, in terms of which one may find a solution of the
model. I refer the interested reader to Ref. \cite{cmp_08} for
computational details.

\section{Magnetic Responses}
\label{sec_IV}

By exploiting the general formulation sketched in the previous
section, I shall now study the magnetic properties of the 1D BEG
model, by restricting the analysis to the case $J<0$. In the
following, I set $J=-1$ and I consider only positive values of
$h$, owing to the symmetry property of the model under the
transformation $h \to -h$ and $S \to-S$. When $K=0$ and
$\Delta=0$, the ground state is purely antiferromagnetic for $0
\le h<2 \vert J \vert$. By varying the magnetic field, the system
undergoes a phase transition to a  paramagnetic state at $h= \pm 2
\vert J \vert$. As a consequence, the magnetization, defined as
\begin{equation}
m=\langle S(i)\rangle=\langle n(i)\rangle-1,
\end{equation}
presents,
\begin{figure}[t]
  \begin{center}
  \begin{tabular}{cc}
  \begin{minipage}[c]{.32\textwidth}
    \begin{center}
   \includegraphics[scale=0.175]{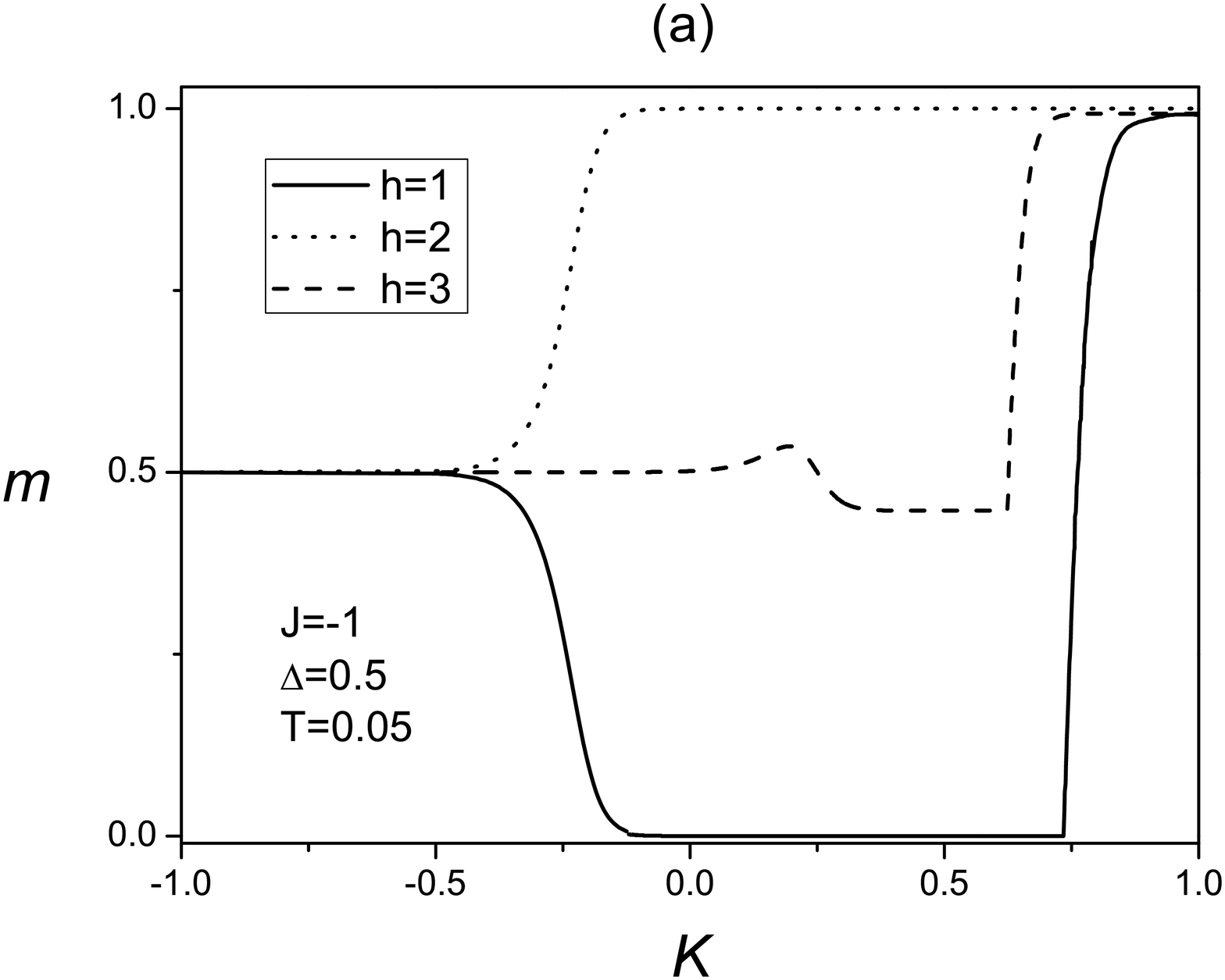}
    \end{center}
  \end{minipage}
    \begin{minipage}[c]{.32\textwidth}
    \begin{center}
   \includegraphics[scale=0.175]{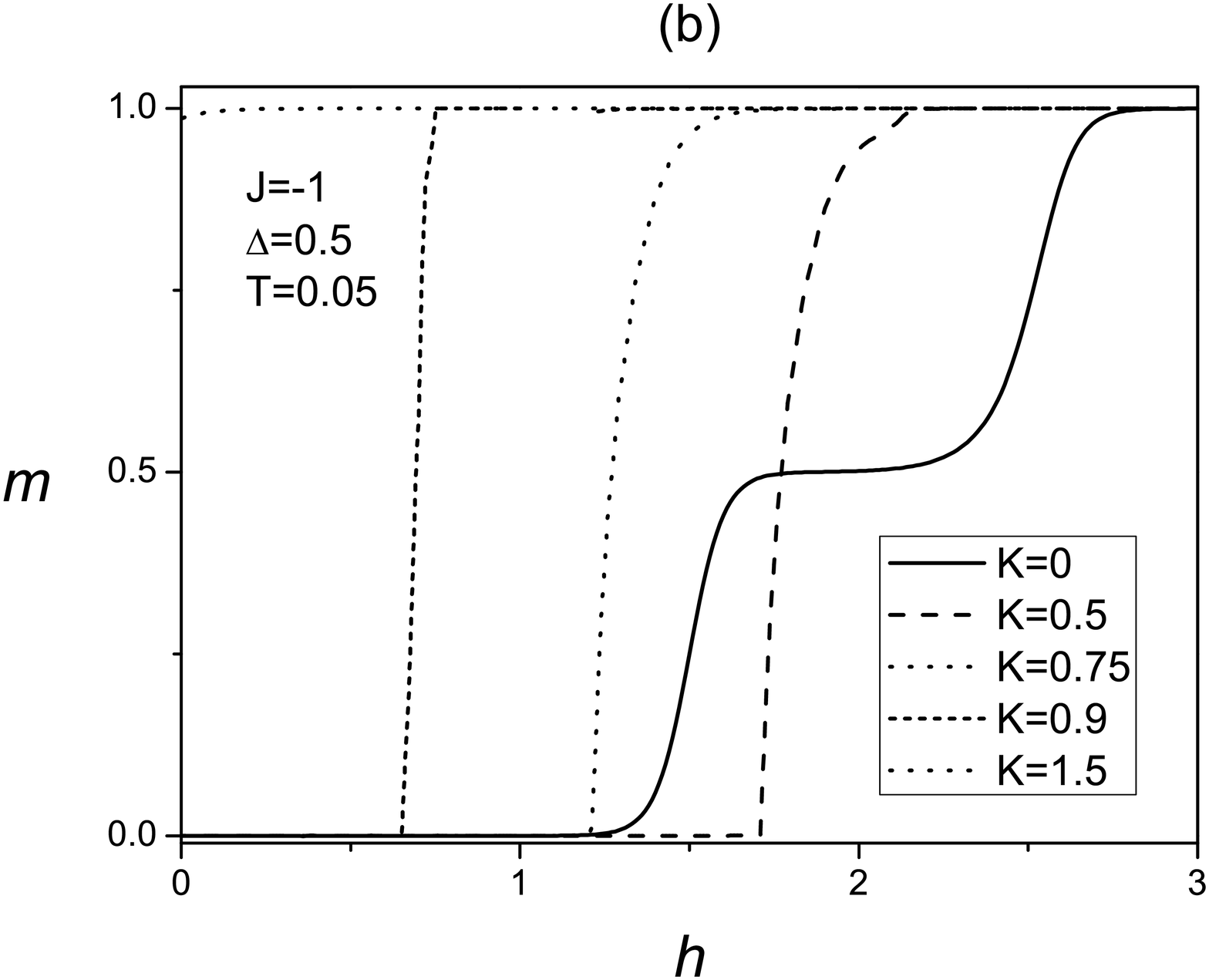}
    \end{center}
  \end{minipage}
 \begin{minipage}[c]{.32\textwidth}
    \begin{center}
   \includegraphics[scale=0.175]{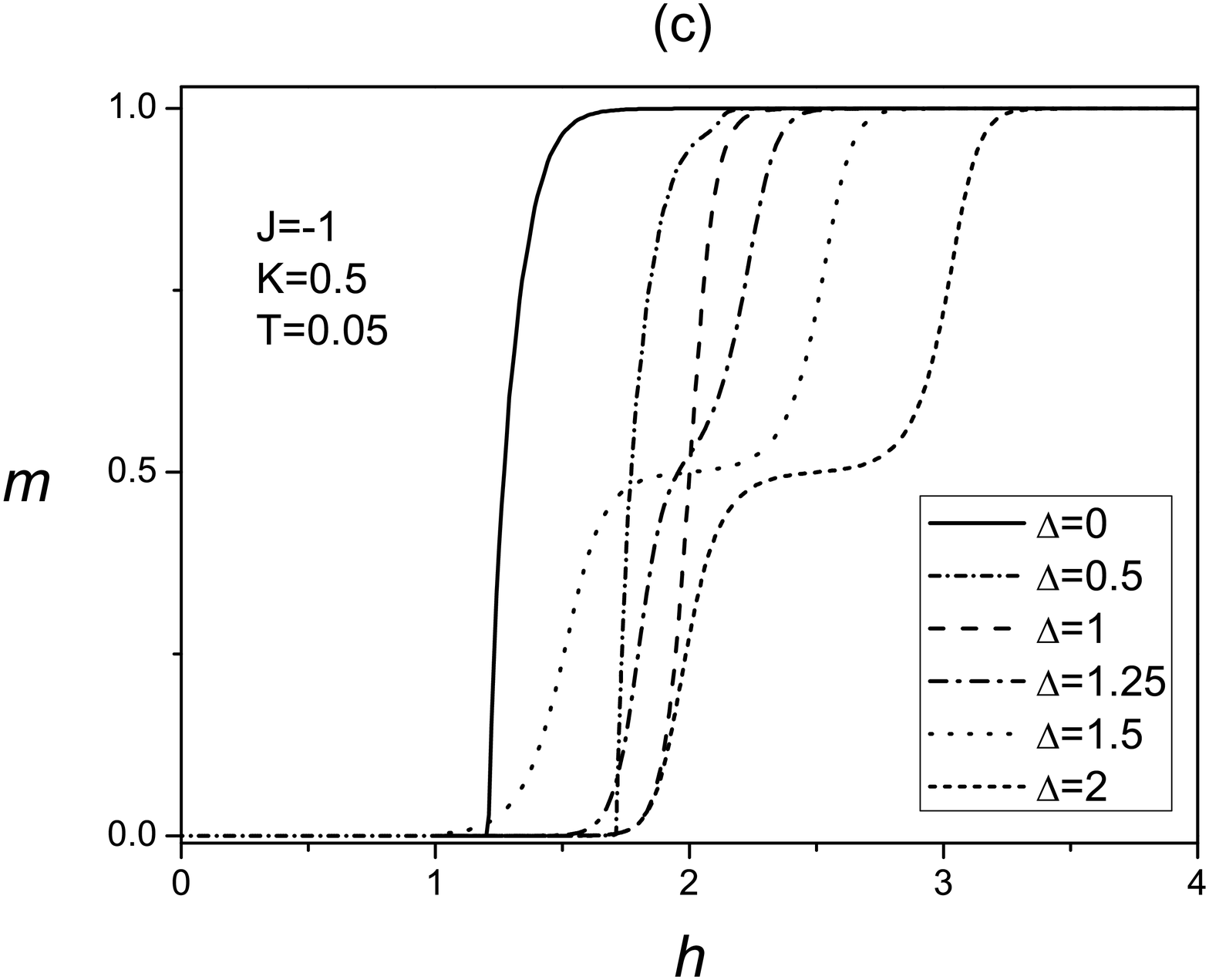}
    \end{center}
  \end{minipage}
  \end{tabular}
  \caption{\label{fig1} (a) The magnetization $m$ as a function of the biquadratic interaction
$K$ for $J=-1$, $T=0.05$, $\Delta=0.5$, and $h=1,2,3$. (b) The
magnetization as a function of the external field $h$ for $J=-1$,
$T=0.05$, $\Delta=0.5$ and positive values of $K$. (c) The
magnetization as a function of the external field for $J=-1$,
$T=0.05$, $K=0.5$ and several  values of $\Delta$.}
\end{center}
\end{figure}
at $T=0$, two plateaus as a function of the external field $h$.
When the anisotropy $\Delta$ is turned on (keeping $K=0$), one
observes three plateaus at $m=0$, $m=1/2$ and $m=1$
\cite{cmp_08,oshikawa}. The intermediate phase between the antiferromagnetic
and paramagnetic ones  has a width depending on $\Delta$, whose
endpoints are denoted by $h_c$ and $h_s$, i.e., starting point of
a nonzero magnetization and saturated field, respectively. For
finite biquadratic interaction, the presence of plateaus in the
magnetization curve dramatically depends on the sign of the
interaction itself and on the value of the anisotropy.

\begin{figure}
  \begin{center}
  \begin{tabular}{cc}
  \begin{minipage}[c]{.475\textwidth}
    \begin{center}
     \includegraphics[scale=0.22]{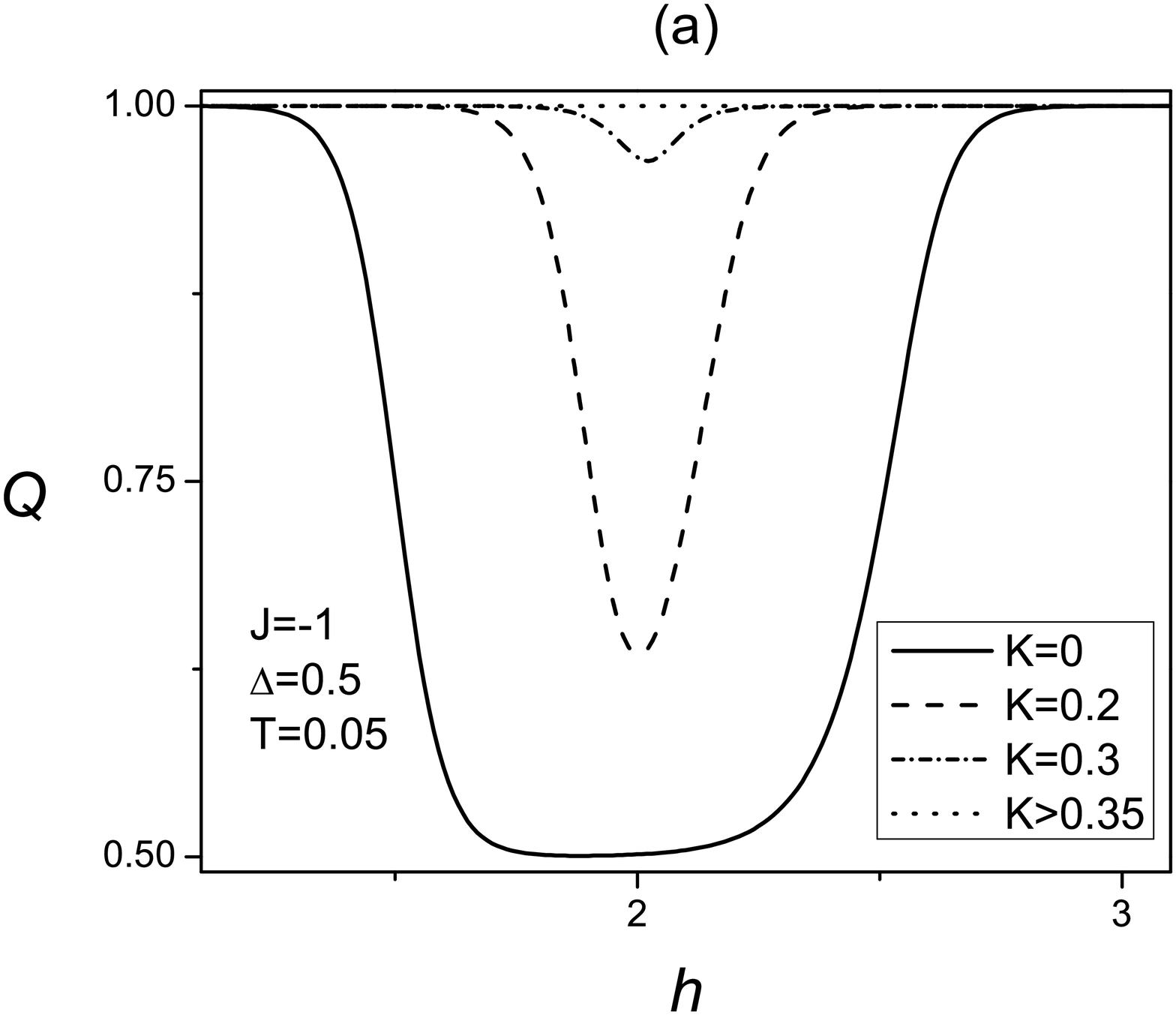}
    \end{center}
  \end{minipage}
  \begin{minipage}[c]{.475\textwidth}
    \begin{center}
    \includegraphics[scale=0.21]{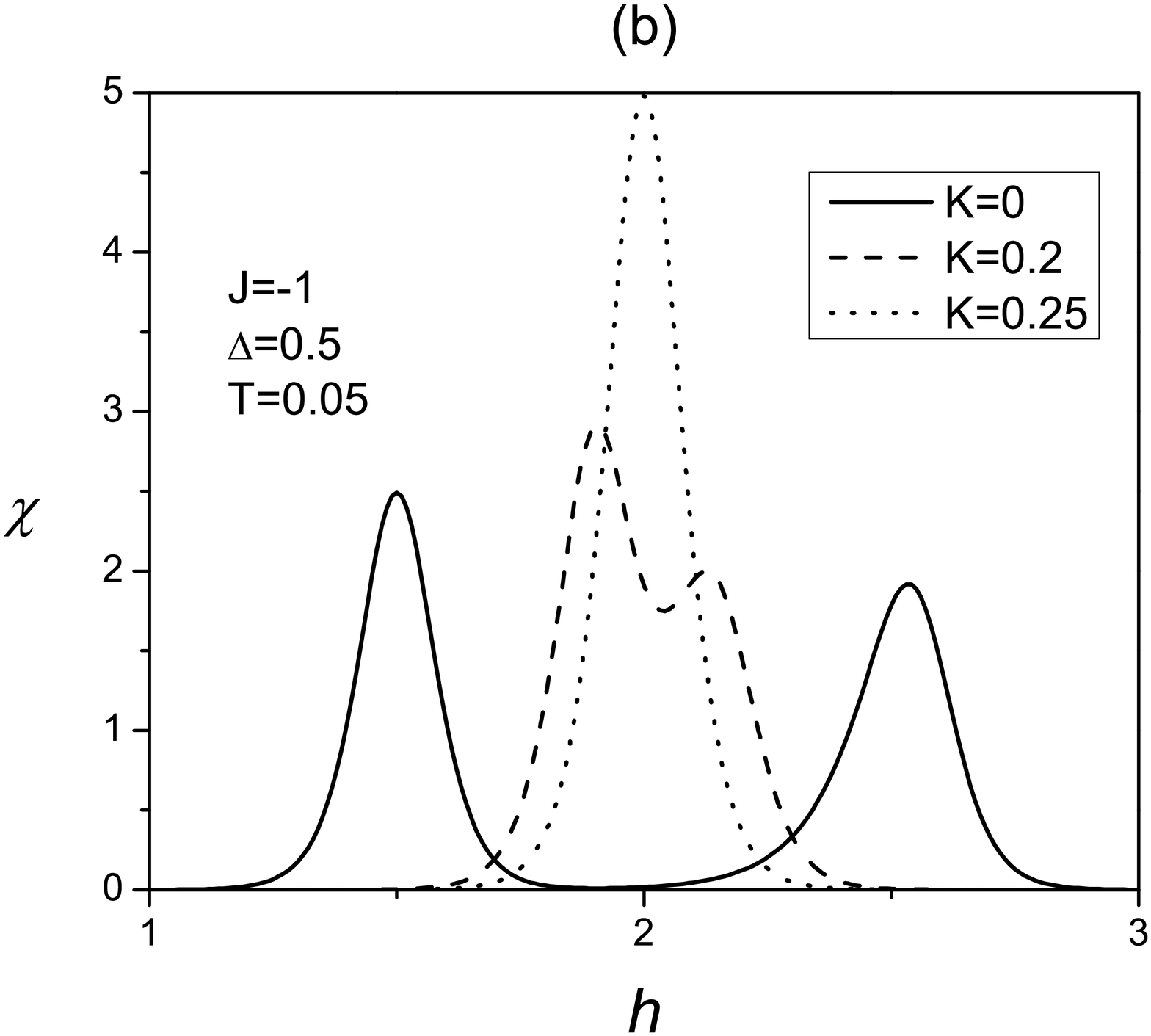}
    \end{center}
  \end{minipage}
  \end{tabular}
   \caption{ \label{fig2} (a) The quadrupolar moment $Q$ as a function of the magnetic
field for $J=-1$, $\Delta=0.5$, $T=0.05$ and positive values of
the biquadratic interaction. (b) The magnetic susceptibility
$\chi$ as a function of the magnetic field for $J=-1$,
$\Delta=0.5$, $T=0.05$ and positive values of the biquadratic
interaction.
   }
\end{center}
\end{figure}

In Fig. \ref{fig1}a, I plot the magnetization as a function of the
biquadratic exchange at $\Delta=0.5$ for values of the magnetic
field belonging to the three different plateaus observed at $K=0$.
For all values of $h$, one observes that, for large positive
biquadratic coupling, all the spins are aligned along the magnetic
field ($m=1$). On the other hand, for large negative biquadratic
coupling, half of the spins are parallel to $h$ and the other half
lies in the transverse plane. In the intermediate region,
according to the strength of the biquadratic coupling, one
observes one, two or three plateaus. The range of this
intermediate region depends on $\Delta$, and for $\Delta \ge 1$
only one plateau at $m=1/2$ is observed for $K<0$. In Fig.
\ref{fig1}b, I plot the magnetization as a function of the
magnetic field at $T=0.05$, $\Delta=0.5$ for different values of
$K$. Starting from $K=0$, by turning on a finite biquadratic
exchange, one observes that the width of the intermediate plateau
shrinks and vanishes ($h_c=h_s$). Further augmenting $K$, one
observes the decrease of $h_s$, which eventually vanishes for
sufficiently large $K$: the spins are all polarized in the $h$
direction as soon as the magnetic field is turned on. However, the
three plateau scenario can be restored by varying the anisotropy,
as it is evident from Fig. \ref{fig1}c, where I plot the
magnetization as a function of the magnetic field at low
temperature for $K=0.5$. For fixed $\Delta$, the width of the
$m=1/2$ plateau decreases by increasing $K$. On the other hand,
for fixed $K$, the width of the intermediate plateau augments by
increasing $\Delta$ in the range $0<\Delta<1$ and becomes
independent of $\Delta$ for $\Delta>1$. For these ranges of the
parameters, the critical field $h_c$ and the saturated field $h_s$
satisfy the laws: $h_c = 2 -\Delta$, $h_s=2+\Delta$. To further
analyze the magnetic behavior of the system, I have studied the
quadrupolar moment $Q$, defined as
\begin{equation}
Q=\langle S^2(i)\rangle.
\end{equation}
In the limit $T \to 0$, and for $K=0$, also this quantity shows
plateaus for $\Delta \ge 0$. In Fig. \ref{fig2}a the quadrupolar
moment $Q$ is plotted as a function of the external magnetic
field, for $\Delta=0.5$, $J=-1$, $T=0.05$ and for various values
of $K$. At $K=0$, $Q$ takes the value $1/2$ in the range $h_c < h
< h_s$, whereas is equal to 1 for all other values of $h$. Upon
increasing the value of the biquadratic coupling, one observes the
breakdown of the intermediate plateau, resulting in a uniform
quadrupolar moment for all values of the magnetic field. For
strong anisotropy, the quadrupolar moment shows the same $S$-shape
curve characterizing the magnetization and one finds $Q=m$.  The
possible breakdown of the magnetic plateaus can also be evidenced
by looking at the peak(s) found in the magnetic susceptibility
$\chi= dm/dh$. As evidenced in Fig.\ref{fig2}b, when plotted as a
function of the magnetic field,  $\chi$ shows two peaks at low
temperature  at $h_c$ and $h_s$ - for $K=0$ and $\Delta=0.5$ -
signalling a step-like behavior of the magnetization. By
increasing $K$ one observes that the two peaks merge into only one
peak: the intermediate plateau at $m=1/2$ has disappeared.

\section{Concluding Remarks}
\label{sec_V}

I have evidenced how the use of the Green's function and equations
of motion formalism leads to the exact solution of the
one-dimensional BEG model. The analysis allows for a comprehensive
study of the model in the whole space of parameters $K$, $J$,
$\Delta$, $h$ and $T$. Here, I have focused on the
antiferromagnetic properties exhibited by the model and I have
shown that the three zero-temperature magnetic plateaus scenario exhibited
by the model when $\Delta >0$ and $K=0$ may break down for
sufficiently large positive and negative biquadratic interaction
$K$. However, a large anisotropy can restore the three plateaus
scenario also for finite $K$. This scenario is endorsed by the
behavior of the quadrupolar moment $Q$ and of the susceptibility
$\chi$.
\section*{Acknowledgments}
I thank  F. Mancini for interesting and fruitful discussions.

\end{document}